# Dynamical Crossover in Soft Colloids below the Overlap Concentration

Xin Li,[1] Luis E. Sánchez-Diáz,[1] Bin Wu,[1] William A. Hamilton,[2] Lionel Porcar,[3] Péter Falus,[3] Yun Liu,[4,5] Changwoo Do,[1] Gregory S. Smith,[1] Takeshi Egami,[6] and Wei-Ren Chen[1,*]

[1]*Biology and Soft Matter Division, Oak Ridge National Laboratory, Oak Ridge, Tennessee 37831, USA*
[2]*Instrument and Source Division, Oak Ridge National Laboratory, Oak Ridge, Tennessee 37831, USA*
[3]*Institut Laue-Langevin, B.P. 156, F-38042 Grenoble CEDEX 9, France*
[4]*Center for Neutron Research, National Institute of Standards and Technology, Gaithersburg, Maryland 20899-6100, USA*
[5]*Department of Chemical Engineering, University of Delaware, Newark, Delaware 19716, USA*
[6]*Department of Materials Science and Engineering and Department of Physics and Astronomy, The University of Tennessee, Knoxville, Tennessee 37996-1508, USA*

The dynamics of soft colloids in solutions is characterized by internal collective motion as well as center-of-mass diffusion. Using neutron scattering we demonstrate that the competition between the relaxation processes associated with these two degrees of freedom results in strong dependence of dynamics and structure on colloid concentration, $c$, even well below the overlap concentration $c^*$. We show that concurrent with increasing inter-particle collisions, substantial structural dehydration and slowing-down of internal dynamics occur before geometrically defined colloidal overlap develops. While previous experiments have shown that the average size of soft colloids changes very little below $c^*$, we find a marked change in both the internal structure and internal dynamics with concentration. The competition between these two relaxation processes gives rise to a new dynamically-defined dilute threshold concentration well below $c^*$.

Soft colloids, such as star polymers, dendrimers, microgels, and some polymer-grafted nanoparticles, are man-made macromolecules consisting of polyelectrolyte arms emanating from the center of molecules. They have attracted considerable attention in part because of their intriguing globular but solvent-accessible colloidal architecture [1-4]. The interest in these systems increased even further because of numerous technological applications owing to the fact that they possess useful features of both polymers and colloids. Consequently they have been studied extensively in the last few decades [5-8].

One key feature that differentiates soft colloids from hard sphere or other densely packed particles is the flexibility of their molecular structure [1-4]. Understanding its influence on their conformation, inter-particle interaction and phase behaviors have been the focus of extensive structural studies of soft colloids. This structural softness also renders an additional degree of freedom to the dynamics of soft colloids in solutions. In addition to the center-of-mass diffusion [9-13], many experiments have demonstrated that the dynamics of soft colloids in solution is characterized by intra-particle collective motions as well [14-21].

The properties of soft colloidal solutions are known to exhibit a strong dependence on the colloid concentration $c$ [1,4]. However, below the geometrically defined overlap concentration $c^*$ existing experimental results show that the size of soft colloids remains constant [22-24]. Does it mean that in the concentration range of $c < c^*$ the conformation and internal dynamics of soft colloids remains the same, independent of the concentration $c$?

We address this question through the neutron scattering study of Poly(amido amine) (PAMAM) dendrimers, an extensively studied soft colloid model system. They are characterized by well-defined hierarchical architectures built from iterative synthesis. In each sequence a new concentric shell consisting of terminal groups is added and leads to the next generation [3,8]. The softness of molecules is directly related to the number of their generation; the degree of softness decreases with an increase in generation [3]. In this work aqueous solutions of generation 4 and 6 (G4 & G6) PAMAM dendrimers are investigated focusing on the competition between the inter- and intra-particle relaxation processes. In spite of extensive dynamical studies of soft colloids [14-21], the interplay between these two dynamical degrees of freedom was never scrutinized for $c < c^*$. We demonstrate that this interplay produces a characteristic concentration dependence of the conformation and internal dynamics even at concentrations well below $c^*$.

The dynamics of the dendrimers in solutions was studied using the neutron spin echo (NSE) spectrometer at IN15, Institut Laue-Langevin. Via encoding the change of neutron velocity with spin precession angle, NSE is a technique for measuring dynamics with energy resolution of $neV$ [14]. One distinct advantage of NSE is its ability to access wide spatial and temporal ranges of dynamics. This unique feature allows simultaneous determination of both types of dynamical processes for dendrimer solutions. From the intermediate scattering functions $F(Q,\tau)$ measured by NSE the collective diffusion coefficients $D_S^{Exp}(Q)$ of G4 and G6

PAMAM dendrimers can be extracted [25] and the results are given in Fig. 1.

For a dilute concentration of $c = 0.02$, the contribution from the inter-dendrimer interaction can be disregarded. As shown in Fig. 1 at low $Q$ the data agrees with the corresponding translational self-diffusion coefficients given by the dotted lines. But significant increases in $D_S^{Exp}(Q)$ are clearly observed in the $Q$ range of $Q > 0.12$ Å$^{-1}$ for G4 and $Q > 0.10$ Å$^{-1}$ for G6. Our small angle neutron scattering (SANS) studies have demonstrated that the intra-dendrimer structural characteristics are indeed reflected within these $Q$ ranges [26-27]. Previous studies of other dendrimer systems have identified the intra-molecular collective motion as the origin of this dynamical enhancement [19-20].

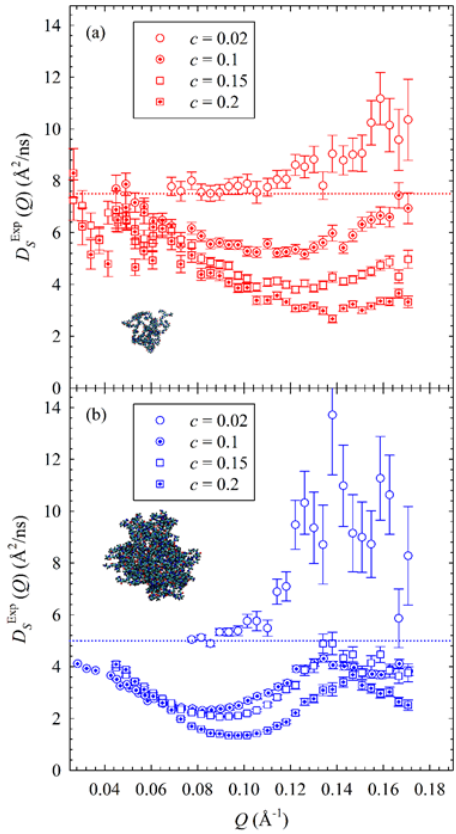

FIG. 1. (color online). The collective diffusion coefficient $D_S^{Exp}(Q)$ of (a) G4 and (b) G6 PAMAM dendrimers as a function of concentration. The dotted lines are the translational self-diffusion coefficients of G4 and G6 PAMAM dendrimers in aqueous solution with $c = 0.02$. In the inset we give the snapshot of G4 and G6 PAMAM dendrimers created by molecular dynamics (MD) simulations [25].

Since the size of soft colloids shows no discernible dependence on concentration when $c < c^*$, [22-24] it is reasonable to assume that all the information of intra-dendrimer dynamics is still contained within the $Q$ range above 0.10 Å$^{-1}$. Indeed upon increasing $c$ to $c^*$, similar variations of $D_S^{Exp}(Q)$ are observed in the $Q$ region relevant to the internal motion. But changes occur also in the lower $Q$ region relevant to larger inter-particle length scales indicating that the measured $D_S^{Exp}(Q)$ contains contributions from both inter- and intra-dendrimer collective motions. Also, within this concentration range a steady establishment of inter-dendrimer interaction is revealed by SANS measurements [25]. Therefore the suppression of $D_S^{Exp}(Q)$ in the whole $Q$ range probed is caused by the increasing inter-dendrimer interaction as well as the hydrodynamic interaction [28-30].

One way to quantify the interplay between the inter- and intra-dendrimer dynamics is via the comparison of their characteristic times. Based on an approximations [25] inspired by analysis of protein dynamics in solutions [31-34], the intra-dendrimer internal motion can be identified separately from the measured $D_S^{Exp}(Q)$ via compartmentalizing the inter-dendrimer contribution [25]. Namely,

$$D_S^{Exp}(Q) \cdot Q^2 = D_S^C(Q) \cdot Q^2 + \Gamma_{intra}(Q) \quad (1)$$

where $D_S^C(Q)$ represent the contribution from the inter-dendrimer collective dynamics and $\Gamma_{intra}(Q)$ is the $Q$-dependent frequency of the intra-dendrimer collective motion. Accordingly we define the characteristic time for inter-dendrimer motion $\tau_{inter}$ as the average collision time between a tagged caged dendrimer and its surrounding neighbors. $\tau_{inter}$ can be determined from the short-time self-diffusion coefficient $D_S^S$, the average inter-dendrimer distance $<L>$ and the size of dendrimer [25]. In addition, after subtracting the contribution from the hydrodynamic interaction, the characteristic time for internal relaxation $\tau_{intra}$ can be obtained from $D_S^{Exp}(Q)$ within the $Q$ range of $Q > 0.12$ Å$^{-1}$ for G4 and $Q > 0.10$ Å$^{-1}$ for G6 [25]. The results of $\tau_{inter}$ and $\tau_{intra}$ as a function of $c$ are given in Fig. 2.

Upon increasing $c$, $\tau_{inter}$ is seen to decrease by several orders of magnitude for both G4 and G6 dendrimers. This observation is a reflection of slowing down of the dendrimer self-motion, characterized by $D_S^S$, and a significant decrease in $<L>$, caused by the increase in concentration [25]. Meanwhile, with the influence of the hydrodynamic interaction, the minima of $D_S^{Exp}(Q)$ provides the lower limit of the dendrimer self-motion and the maxima provides the upper limit of the total dynamics including the internal motion. Therefore, the difference between these two extremes gives the upper limit for the frequency of internal relaxation. Results of this qualitative estimation indicate a discernible increase of $\tau_{intra}$ upon increasing $c$ below $c^*$. A quantitative analysis demonstrated in [25] suggests the increase in dendrimer concentration from $c = 0.02$ to $0.2$ renders an increase in $\tau_{intra}$ by a factor of 4 and 40 for both G4 and G6 dendrimers. Although whether the NSE-measured internal collective motion of dendrimer is due to shape fluctuation or density fluctuation remains a subject of ongoing scientific discussion [14], the increase in $\tau_{intra}$ below $c^*$ clearly indicates that this internal dynamics are progressively restricted and decelerated by the increase in the frequency of physical contact with its neighbors due to the concentration effect. From the SANS data analysis presented in [25], it is reasonable to assume that the equilibrium conformations of both G4 and G6 dendrimers at $c = 0.02$ are free from the influence of inter-dendrimer interaction. The increasing $\tau_{intra}$ indicates that even below $c^*$ dendrimers are indeed restrained from fully relaxing to their original conformation at $c = 0.02$ due to the progressively shortened dendrimer-dendrimer collision time interval.

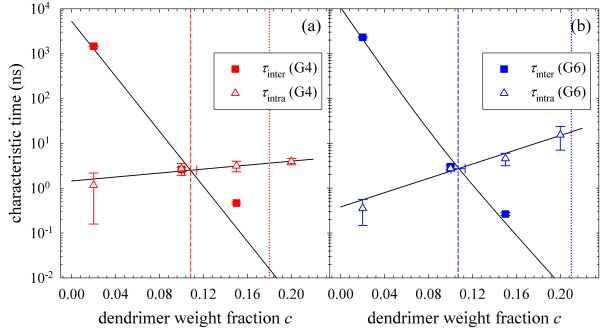

FIG. 2. (color online). The average inter-dendrimer collision time $\tau_{inter}$ (open circles) and internal relaxation time $\tau_{intra}$ (filled circles) for (a) G4 and (b) G6 PAMAM dendrimers in water as a function of concentration $c$. The relative error for $\tau_{intra}$ is shown in the figure, and for $\tau_{inter}$ it is smaller than the symbol size in logarithm scale. Details of error calculation are given in [25]. The dotted and dashed lines respectively give the concentrations of $c^*$ and $c_D^*$ of G4 and G6 PAMAM dendrimers. $c^*$ are estimated according to [35]. Assuming both $\tau_{inter}$ and $\tau_{intra}$ evolve exponentially with $c$, $c_D^*$ can be defined by the intersection of two solid lines. The errors of $c_D^*$ are also presented.

We also explored the conformational changes associated with the sluggish internal motions caused by the dynamical interplay. The conformation of dendrimer was shown to be closely related to the invasive water [36-37]. It has been demonstrated that the packing of invasive water is looser than that of the bulk state of water [25,37]. By including the coherent scattering contribution from the water, one may extract the concentration dependence of dendrimer conformation below $c^*$ from SANS experiments. The total scattering power of a single dendrimer, which we denote as $P(0)$, is a function of bound scattering lengths of the constituent atoms of polymer, the average scattering length density (SLD) of water, the packing density of invasive water and the volume of the intra-dendrimer cavities. By analyzing $P(0)$ [25], one can evaluate the conformational evolution of dendrimer SLD from the variation of a single dendrimer $<\rho>$, which takes the following expression

$$\langle \rho \rangle = \frac{b_{polymer}}{v_{polymer} + (1 - v_{water} \cdot h) \cdot v_{cavity}} \quad (2)$$

$b_{polymer}$ and $v_{polymer}$ in the right hand side (RHS) of Eq. (2) are the summation of the bound scattering lengths and the volume of the constituent atoms of the polymer components of dendrimer, respectively, $h$ is the number density of invasive water molecules, $v_{water}$ is the molecular volume of water in its bulk state, and $v_{cavity}$ is the total volume of the intra-molecular cavities. We show the $<\rho>$ for G4 and G6 PAMAM dendrimer solutions studied in this work in Fig. 3.

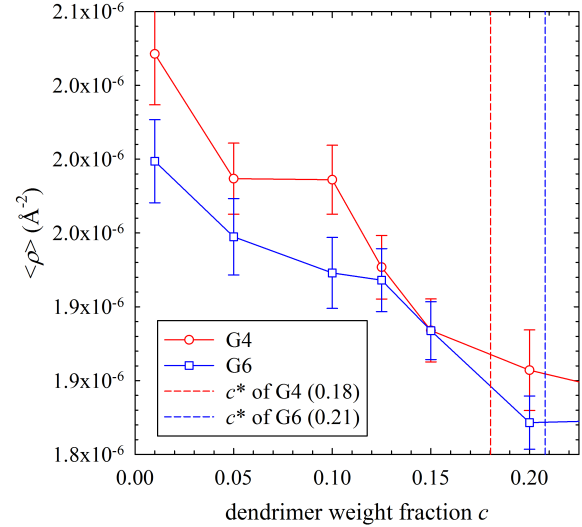

FIG. 3. (color online). The average SLD of a single dendrimer $<\rho>$ for G4 and G6 PAMAM dendrimer as a function of dendrimer weight fraction in solutions. The lines are used as a guide to eyes.

The most important observation in Fig. 3 is that, for both dendrimer solutions, $<\rho>$ is a decreasing function of $c$ within the concentration range of $c < c^*$ [25]. While explicitly separating the individual contribution of each physical quantity in the RHS of Eq. (1) to the evolution of $<\rho>$ is not possible with our current methodology, a qualitative picture of the conformational evolution below $c^*$ can certainly be deduced from the following argument. Provided that both $b_{polymer}$ and $v_{polymer}$ do not exhibit dependence on $c$, the decreasing trend of $<\rho>$ in Fig. 3 suggests that the magnitude of $(1 - v_{water} \cdot h) \cdot v_{cavity}$ must increase upon increasing $c$. Meanwhile, since the excluded volume effect can only cause a contraction of intra-molecular voids, it is clear that $v_{cavity}$ is not an increasing function of $c$. As a result, $h$, the parameter quantifying the packing of invasive water, must decrease to ensure $(1 - v_{water} \cdot h) \cdot v_{cavity}$ is an increasing function of $c$. Therefore, as a structural reflection of the progressively sluggish internal dynamics, steady dehydration of a single dendrimer caused by the increase in $c$ well below $c^*$ and resultant increase in the frequency of inter-particle collision can be concluded unambiguously. It is instructive to compare the observed evolution of $<\rho>$ with $c$ with existing results of conformational study of soft colloids below $c^*$. A static conformational picture of the soft colloid below $c^*$ has been proposed based on the invariance of radius of gyration ($R_G$) observed by scattering experiments [22-24]. According to its mathematical definition [38], $R_G$ reflects only the integrated, thus coarse-grained, information of the intra-particle density profile. It is therefore possible that the detailed internal structural variation presented here is smeared out during the integration process and therefore may not be explicitly reflected in $R_G$.

Comparing the evolution of $\tau_{intra}$ presented in Fig. 2 with that for $<\rho>$ given in Fig. 3, one sees that $\tau_{intra}$ for G6 is discernibly longer than that of G4 at some certain concentrations. Since G4 in water is proven to be softer than G6 in terms of SLD distribution, this observation is somehow counterintuitive. However, experimentally we have demonstrated that the intra-molecular hydration level is higher in the lower-generation dendrimer [37]. Using quasielastic neutron scattering (QENS) and atomistic MD simulation, we also showed that the internal relaxation of polymer components of a dendrimer was indeed driven by their interaction with invasive water [39]: More invasive water results in faster relaxation. Therefore it is consistent to see the connection between the slowing down of internal relaxation and the molecular dehydration, as well as their generational dependence. One may also note that the results shown in Figs. 2 and 3 are characterized by very similar qualitative features. This observation suggests that the conformational and dynamical evolution below $c^*$ may be a universal feature of general soft colloids. Moreover, because both $\tau_{intra}$ and $\tau_{inter}$ showed in Fig. 2 evolve continuously within the range of $c < c^*$, the intersection of evolving $\tau_{intra}$ and $\tau_{inter}$ must exist which is considerably less than the geometrically-defined $c^*$. Assuming both $\tau_{intra}$ and $\tau_{inter}$ evolve exponentially as a function of $c$, $c_D^*$ of G4 and G6 dendrimers are found to be around 10 wt%. The physical meaning of $c_D^*$ is that it marks a dynamically-defined crossover. Below $c_D^*$ the inter-dendrimer collision time $\tau_{inter}$ is longer than the colloidal relaxation time $\tau_{intra}$, so that a colloid particle recovers its original shape after collision. Beyond $c_D^*$, however, colloids have no time to recover, and retains the deformed state induced by inter-particle collision. Therefore these two degrees of freedom are coupled only when $c > c_D^*$.

In conclusion, we investigated the general conformational and dynamical behavior of soft colloid by means of neutron scattering using dendrimers as an example of soft colloid. Unlike their hard relatives, soft colloids exhibit significant internal dynamics in addition to the center-of-mass diffusive phenomena. Even well below $c^*$, we find that inter-particle interactions begin to result in persistent structural dehydration and sluggish internal motions. The origin of this unexpected dynamical and structural evolutions below $c^*$ is the competition between the inter- and intra-particle relaxation mechanisms. In contrast to the geometrically defined $c^*$, a dynamically defined dilute threshold concentration $c_D^*$ is manifested by the crossover of the relaxation times of the two competing relaxation mechanisms.

We thank Profs. Phil Pincus, Magdaleno Medina-Noyola, David Callaway, Zimei Bu, and Kostas Karatasos for helpful discussions. This work was supported by the U.S. Department of Energy, Office of Basic Energy Sciences, Materials Sciences and Engineering Division. This Research at the SNS at Oak Ridge National Laboratory was sponsored by the Scientific User Facilities Division, Office of Basic Energy Sciences, U.S. Department of Energy. We greatly appreciate the D22 SANS and IN15 NSE beamtime from the Institut Laue-Langevin.

## Supplementary Material

## SI. Sample Preparation

The G4 and G6 PAMAM dendrimers used in this work were purchased from Dendritech Inc., Midland, MI, USA. Deuterium chloride (catalog number DLM-54-25) and deuterium oxide $D_2O$ (catalog number DLM-6-10X1) were obtained from Cambridge Isotope Laboratories, Inc., Andover, MA, USA. The samples were prepared by dissolving PAMAM dendrimer in solutions of $D_2O$ according to that targeted dendrimer weight fraction $c$. Solvents used for the SANS contrast matching experiments were prepared by mixing a predetermined amount of $D_2O$ and de-ionized water (purified from Millipore system) with molar ratios of $D_2O$ to $H_2O$ at 100:0, 90:10, 80:20, 70:30, and 60:40.

In their own activities as scientific institutions, NIST and ORNL use many different materials, products, types of equipment, and services. However, NIST and ORNL do not approve, recommend, or endorse any product or proprietary material.

## SII. Neutron Spin Echo Experiment

The neutron spin echo (NSE) measurements were carried out at the IN15 at ILL, Grenoble, at wavelengths from 6 Å up to 15 Å in a temperature controlled environment at $20^{\circ}C$. The G4 and G6 dendrimers were dissolved in $D_2O$ with concentrations $c$ of 0.02, 0.1, 0.15 and 0.2. Judging from the size of G4 dendrimer ($R_G$ = 20 Å determined from SANS data analysis), the probed $Q$ range was set to be 0.03 Å$^{-1}$ < $Q$ < 0.18 Å$^{-1}$. Examples of the measured intermediate scattering function $F(Q,\tau)$ are given in Figures S3 for G4 and G6 dendrimers, respectively, and the measured diffusion coefficient $D_S^{Exp}(Q)$ was determined from the decay constant of $F(Q,\tau)$ via Eq. (S8).

## SIII. Neutron Spin Echo Data Analysis

The short-time collective diffusion coefficient containing the hydrodynamic contribution can be expressed as

$$D_S^C(Q) = D_0 \frac{H(Q)}{S(Q)} \quad (S1)$$

where $D_0$ is the diffusion coefficient of the infinite dilute limit, $H(Q)$ is the hydrodynamic function which depends on the hydrodynamic interaction, inter-particle effective interaction, and volume fraction, and $S(Q)$ is the structure factor. It is clear that the information of $D_S^C(Q)$ is contained in the experimentally measured $D_S^{Exp}(Q)$ given in Figure 1 of this letter. The following approach is developed to properly identify $D_S^C(Q)$. First the corresponding SANS $I(Q)$ was analyzed based on the previously developed factorization approximation [S1] which is valid in the concentration range of $0 < c < c^*$ studied in this work according to [S2]. The previously proposed Gaussian potential function was used to model the inter-dendrimer interaction, and the $S(Q)$ was obtained via solving the Ornstein-Zernike integral equation (OZ) with the Percus–Yevick (PY) closure. [S3] To achieve a satisfactory agreement between experiment and theory, we found that the repulsive strength of the Gaussian potential $V(r=0)$ needs to be treated as a fitting parameter rather than fixed as a constant. Its values obtained at different concentration c are given in Table SI.

**Table SI.** Physical parameters for G4 and G6 solutions measured in NSE experiments

| Dendrimer sample | Weight percentage c | $Q^*$ (Å$^{-1}$) | $\phi_{eff}$ | $<L>$ (Å) | $V(r = 0)$ ($k_BT$) |
|---|---|---|---|---|---|
| G4 | 0.02 | N/A | N/A | 107.9374 | N/A |
| G4 | 0.1 | 0.11157 | 0.22503 | 59.9069 | 4.6091 |
| G4 | 0.15 | 0.11609 | 0.28891 | 56.9973 | 4.8863 |
| G4 | 0.2 | 0.12914 | 0.37252 | 51.3478 | 5.7062 |
| G6 | 0.02 | N/A | N/A | 141.804 | N/A |
| G6 | 0.1 | 0.078487 | 0.27033 | 84.2141 | 5.2748 |
| G6 | 0.15 | 0.081154 | 0.33851 | 81.7141 | 5.601 |
| G6 | 0.2 | 0.085131 | 0.36739 | 78.7154 | 6.5276 |

In practical experiments for any particle whose internal collective motion is negligible such as the densely packed colloid, its $D_S^{Exp}(Q)$ is indeed equal to $D_S^C(Q)$. Therefore, the short-time self-diffusion coefficient $D_S^S$ is usually obtained from the measurement of $D_S^{Exp}(Q)$ in the high $Q$ region, namely

$$D_S^S = \lim_{Qd>10} D_S^{Exp}(Q) \tag{S2}$$

where *d* is the particle diameter. However, for soft colloid this approach to obtain $D_S^S$ is severely compounded by the intra-particle internal dynamics, as demonstrated in Figure 1.

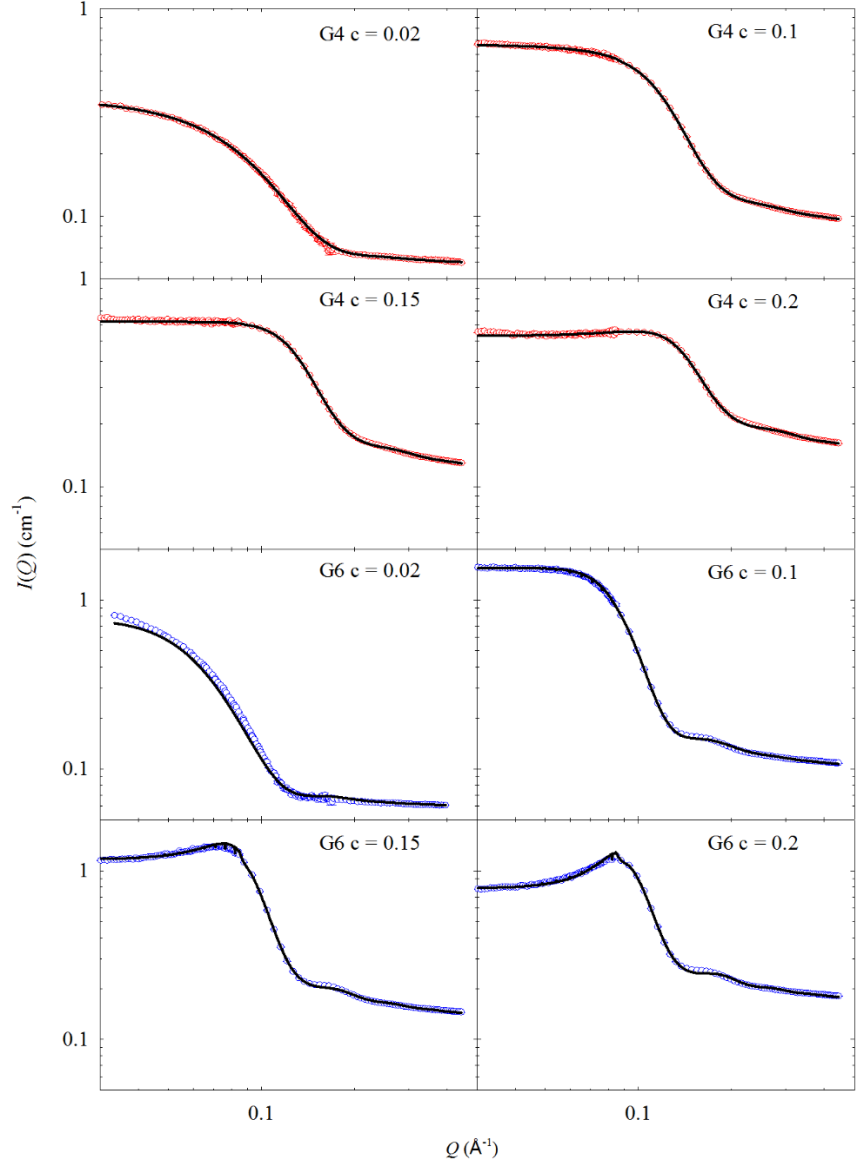

**FIG. S1.** (color online). The scattering cross section *I(Q)* obtained for the G4 and G6 PAMAM dendrimer solutions at the same concentration for NSE experiments. The symbols are the experimental results and the black lines are the model fitting data. The statistical error is smaller than the size of symbol.

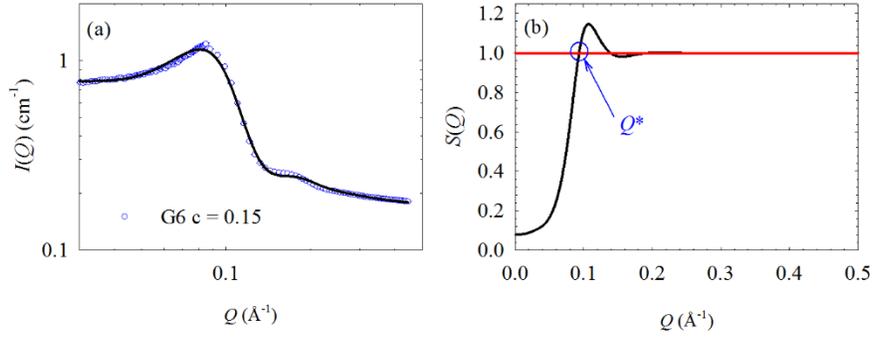

**FIG. S2.** (color online). An example to demonstrate the location of $Q^*$ of $S(Q)$.

To bypass the complication caused by the internal collective motion, the $D_S^S$ of dendrimer solutions was obtained via an alternative approach first proposed by Pusey [S4], which has recently been validated by a theoretical and computational study [S5]. In this approach, $D_S^S$ can be obtained via

$$D_S^S = D_S^{Exp}(Q^*) \tag{S3}$$

where $S(Q^*) = 1$ as shown in Figure S2(b). This approach allows us to obtain the value of $D_S^S$ in the much lower $Q$ region which is beyond the length scale relevant to the internal dynamics. Examples of SANS $I(Q)$ along with the $S(Q)$ and the value of $Q^*$ is given in Figure S2. The value of $D_S^S$ at each dendrimer weight fraction $c$ is given in Table SI.

From $Q_{max}$, the peak position of the $S(Q)$, the average inter-dendrimer distance $<L>$ is estimated via the following expression [S6]

$$\langle L \rangle = \frac{2.5\pi}{Q_{max}} \tag{S4}$$

The average inter-dendrimer collision time $\tau_{inter}$ is therefore determined from

$$\tau_{inter} = \frac{[\langle L \rangle - 2 \times (1.5 R_G)]^2}{D_S^S} \Big/ P \tag{S5}$$

where $P$ is the collision probability given by

$$P = \frac{4\pi \cdot \pi R_G^2}{4\pi \langle L \rangle^2} \tag{S6}$$

In Eqn. (S5) the molecular boundary is chosen to be 1.5 $R_G$ based on the definition of $c^*$ according to [S2]. It is subtracted from the inter-dendrimer distance to get the exact separation between the boundaries of neighboring dendrimer molecules. The collision frequency is corrected by the cross section of dendrimer molecules along the moving direction. The factor $4\pi$ present in the numerator of Eqn. (S6) represents the coordination number generally found for disordered systems like liquids and glasses. [S7]

Through Eqns. (S4 - S6) and using standard deviation, the error of $\tau_{inter}$ can be estimated from the following equation:

$$\left(\frac{\Delta\tau_{inter}}{\tau_{inter}}\right)^2 = \left(\frac{2\Delta(\langle L \rangle - 3R_G)}{\langle L \rangle - 3R_G}\right)^2 + \left(\frac{\Delta D_S^S}{D_S^S}\right)^2 + \left(\frac{\Delta P}{P}\right)^2$$

$$= \left(\frac{2\Delta(\langle L \rangle - 3R_G)}{\langle L \rangle - 3R_G}\right)^2 + \left(\frac{\Delta D_S^S}{D_S^S}\right)^2 + \left(\frac{2\Delta R_G}{R_G}\right)^2 + \left(\frac{2\Delta\langle L \rangle}{\langle L \rangle}\right)^2$$

$$= \left(\frac{2\Delta\left(\frac{2.5\pi}{Q_{max}} - 3R_G\right)}{\frac{2.5\pi}{Q_{max}} - 3R_G}\right)^2 + \left(\frac{\Delta D_S^S}{D_S^S}\right)^2 + \left(\frac{2\Delta R_G}{R_G}\right)^2 + \left(\frac{2\Delta Q_{max}}{Q_{max}}\right)^2 \quad (S7)$$

$$= \left(\frac{2\sqrt{\Delta Q_{max}^2 \cdot \left(\frac{2.5\pi}{Q_{max}^2}\right)^2 + 9\Delta R_G^2}}{\frac{2.5\pi}{Q_{max}} - 3R_G}\right)^2 + \left(\frac{\Delta D_S^S}{D_S^S}\right)^2 + \left(\frac{2\Delta R_G}{R_G}\right)^2 + \left(\frac{2\Delta Q_{max}}{Q_{max}}\right)^2$$

where $\Delta D_S^S$, $\Delta R_G$, and $\Delta Q_{max}$ are the statistical errors of $D_S^S$, $R_G$, and $Q_{max}$ obtained respectively from the data fitting. For the dilute cases with the concentration $c = 0.02$, it is assumed that there is no inter-dendrimer structure. Therefore, $\Delta\tau_{inter}$ is calculated using the second line of Eqn. (S7) and the fourth term in the right hand side is zero. Since there is no structure factor at this concentration, the distance $<L>$ is calculated as the cubic root of the inverse of the number density. The values of $D_S^S$, $R_G$, $Q_{max}$ and their statistical errors are given in Table SII.

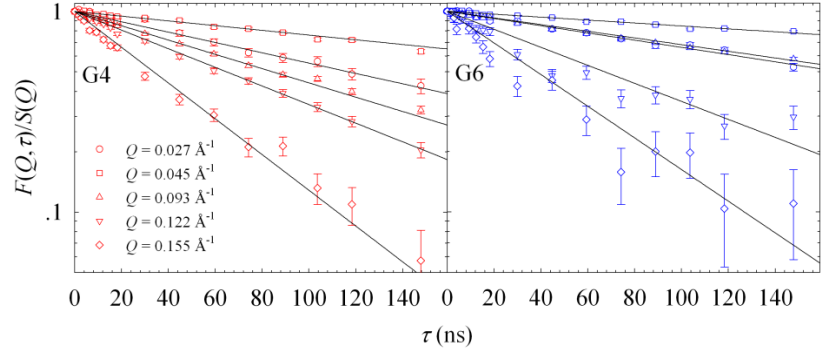

**FIG. S3.** (color online). An example of the measured intermediate scattering function $F(Q,\tau)$ for G4 and G6 solutions with concentration $c = 0.15$.

**Table SII.** Physical parameters and their statistic errors for G4 and G6 solutions measured in NSE experiments

| Dendrimer sample | Weight percentage $c$ | $R_G$ (Å) | $\Delta R_G$ (Å) | $Q_{max}$ (Å$^{-1}$) | $\Delta Q_{max}$ (Å$^{-1}$) | $D_S^S$ (Å$^2$/ns) | $\Delta D_S^S$ (Å$^2$/ns) |
|---|---|---|---|---|---|---|---|
| G4 | 0.02 | 18.357 | 0.428 | N/A | N/A | 7.5 ($D_0$) | 0.610 |
| G4 | 0.1 | 17.611 | 0.352 | 0.11157 | 0.002 | 5.442 | 0.134 |
| G4 | 0.15 | 17.301 | 0.346 | 0.11609 | 0.001 | 4.066 | 0.220 |
| G4 | 0.2 | 17.182 | 0.533 | 0.12914 | 0.002 | 3.158 | 0.314 |
| G6 | 0.02 | 27.295 | 0.994 | N/A | N/A | 5.0 ($D_0$) | 0.090 |
| G6 | 0.1 | 26.703 | 0.538 | 0.07848 | 0.023 | 2.366 | 0.023 |
| G6 | 0.15 | 26.315 | 0.510 | 0.08115 | 0.021 | 2.173 | 0.021 |
| G6 | 0.2 | 26.255 | 0.662 | 0.08513 | 0.018 | 1.448 | 0.062 |

With the effective hard sphere volume fraction $\phi_{eff}$ obtained from our SANS model fitting as the input, the hydrodynamic function $H(Q)$ at different dendrimer concentration can be obtained from the theoretical framework proposed by Snook et al.. [S8] Within the concentration range studied by NSE, it has been shown [S9] that the empirical expression of $H(Q)$ by Snook et al. is in a good agreement with that obtained by a more rigorous method by Beenakker and P. Mazur [S10-S12]. As shown in Figure S6, fairly good agreement between the experiment and theory is seen within the $Q$ range relevant to the inter-dendrimer length scale ($Q < 0.1$ Å$^{-1}$). We assume that the measured short-time diffusion coefficient $D_S^{\exp}(Q)$ within the $Q$ range of 0.12 Å$^{-1}$ $< Q <$ 0.18 Å$^{-1}$ is a combination of $D_S^C(Q)$ contributed by the inter-dendrimer collective dynamics and that originating from the intra-dendrimer collective motion. Explicitly, if the measured intermediate scattering function takes the following expression

$$F_{\exp}(Q,\tau) = S(Q)\exp\left(-D_S^{Exp}(Q) \cdot Q^2 \cdot \tau\right) \tag{S8}$$

and

$$F_{hyd}(Q,\tau) = S(Q)\exp\left(-D_S^C(Q) \cdot Q^2 \cdot \tau\right) \tag{S9}$$

The measured short-time diffusion coefficient $D_S^{Exp}(Q)$ can be expressed by the following equation [S20-S23]

$$D_S^{Exp}(Q) = \frac{k_B T}{Q^2} \times \frac{\sum_{jl}\langle b_j b_l \left(Q \cdot H^T \cdot Q + L_j \cdot H_{jl}^R \cdot L_l + M(Q)\right) e^{iQ \cdot (r_j - r_l)}\rangle}{\sum_{jl}\langle b_j b_l e^{iQ \cdot (r_j - r_l)}\rangle} \tag{S10}$$

where $b_j$ is the coherent scattering length of the subunit $j$, $r_j$ the coordinate for that subunit, $L_j = Q \times r_j$ the angular momentum of that coordinate, $H^T$ the translational mobility tensor, $H^R$ the rotational mobility tensor, and $k_B T$ the thermal energy unit. Since the microscopic mechanism of the internal collective motion still remains unknown, we use a $Q$-dependent value $M(Q)$ to describe its contribution on the measured dynamics. The brackets $\langle\rangle$ denote the orientational average over $Q$.

The translational and rotational diffusion can be estimated using Stokes-Einstein equation as follows

$$D_0^T = \frac{k_B T}{6\pi \eta_0 R} \tag{S11}$$

and

$$D_0^R = \frac{k_B T}{8\pi \eta_0 R^3} \tag{S12}$$

where $D_0^T$ and $D_0^R$ are the translational and rotational diffusion coefficients at the dilute limit, $\eta_0$ the solvent viscosity, and $R$ the colloid radius. The ratio between them gives

$$\frac{D_0^R}{D_0^T Q^2} = \frac{3}{4(QR)^2} \tag{S13}$$

Since the rotation cannot be in a length scale larger than the particle size, which means

$$Qd = 2QR > 2\pi \tag{S14}$$

where *d* is the diameter of the colloid particle. Substituting Eqn. (S14) into (S13) gives us that, in this *Q* range, the ratio of the contribution of rotational diffusion to that of translational diffusion is much less than 1. Therefore, the contribution from rotational diffusion is negligible. Eqn. (S10) can be further simplified and compartmentalized as

$$D_S^{Exp}(Q) = \frac{k_B T}{Q^2} \times \frac{\sum_{jl}\langle b_j b_l (Q \cdot H^T \cdot Q + M(Q)) e^{iQ \cdot (r_j - r_l)}\rangle}{\sum_{jl}\langle b_j b_l e^{iQ \cdot (r_j - r_l)}\rangle}$$

$$= \frac{k_B T}{Q^2} \times \left[\frac{\sum_{jl}\langle b_j b_l (Q \cdot H^T \cdot Q) e^{iQ \cdot (r_j - r_l)}\rangle}{\sum_{jl}\langle b_j b_l e^{iQ \cdot (r_j - r_l)}\rangle} + \frac{\sum_{jl}\langle b_j b_l M(Q) e^{iQ \cdot (r_j - r_l)}\rangle}{\sum_{jl}\langle b_j b_l e^{iQ \cdot (r_j - r_l)}\rangle}\right] \quad (S15)$$

Namely

$$D_S^{Exp}(Q)Q^2 = D_S^C(Q)Q^2 + \Gamma_{intra}(Q) \quad (S16)$$

Where

$$D_S^C(Q) = \frac{k_B T}{Q^2} \times \frac{\sum_{jl}\langle b_j b_l (Q \cdot H^T \cdot Q) e^{iQ \cdot (r_j - r_l)}\rangle}{\sum_{jl}\langle b_j b_l e^{iQ \cdot (r_j - r_l)}\rangle} \quad (S17)$$

and

$$\Gamma_{intra}(Q) = k_B T \times \frac{\sum_{jl}\langle b_j b_l M(Q) e^{iQ \cdot (r_j - r_l)}\rangle}{\sum_{jl}\langle b_j b_l e^{iQ \cdot (r_j - r_l)}\rangle} \quad (S18)$$

$\Gamma_{intra}(Q)$ is the *Q*-dependent relaxation rate of the intra-dendrimer collective motion. Its relationship with $\tau_{intra}(Q)$ is

$$\Gamma_{intra}(Q)^{-1} = \tau_{intra}(Q) \quad (S19)$$

To represent the trend of the evolution for intra-dendrimer collective motion, we select the average internal relaxation time in the *Q* range where the significant increases of $D_S^{Exp}(Q)$ over $D_S^C(Q)$ are clearly observed.

$$\tau_{intra} = \frac{\int \tau_{intra}(Q) dQ}{\int dQ} \quad (S20)$$

The error of $\tau_{intra}$ originates from the statistical error in the NSE measurement via Eqns. (S15 – S20). Within the range of $Q > \frac{2.5\pi}{2 \times 1.5 R_G}$ which is corresponding to the interior space of a dendrimer molecule, it is found that

$$\frac{\Delta \tau_{intra}(Q)}{\tau_{intra}(Q)} = \frac{\Delta \Gamma_{intra}(Q)}{\Gamma_{intra}(Q)} = \frac{\Delta\left(D_S^{Exp}(Q) - D_S^C(Q)\right)}{D_S^{Exp}(Q) - D_S^C(Q)} = \frac{\sqrt{\left(\Delta D_S^{Exp}(Q)\right)^2 + \left(\Delta D_S^C(Q)\right)^2}}{D_S^{Exp}(Q) - D_S^C(Q)} \quad (S21)$$

The error of the average internal relaxation time $\tau_{\text{intra}}$ can be estimated using the error propagation of an arithmetic average. The values of $\tau_{inter}$, $\tau_{intra}$ and their statistical errors are given in Table SIII.

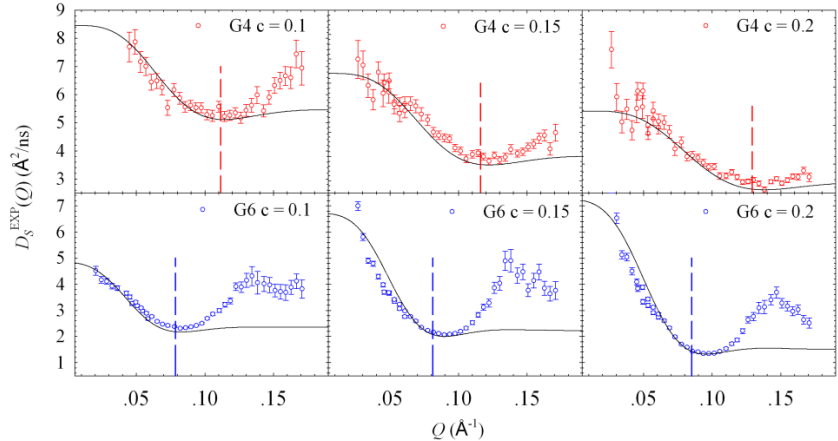

**FIG. S4.** (color online). The measured diffusion coefficient $D_S^{\text{Exp}}(Q)$ for G4 and G6 at the concentrations of $c$ = 0.1, 0.15, and 0.2. The black solid lines are the theoretically calculated $D_S^C(Q)$ using the framework from Snook, and the dashed lines indicate the position of $Q^*$ determined from SANS data analysis for each case.

**Table SIII.** Physical parameters for G4 and G6 solutions measured in NSE experiments

| Dendrimer sample | Weight percentage $c$ | $\tau_{inter}$ (ns) | $\Delta\tau_{inter}$ (ns) | $\tau_{intra}$ (ns) | $\Delta\tau_{intra}$ (ns) |
|---|---|---|---|---|---|
| G4 | 0.02 | 1458.316 | 187.991 | 1.1693 | 1.0104 |
| G4 | 0.1 | 2.6270 | 0.2807 | 2.7312 | 0.8206 |
| G4 | 0.15 | 0.4671 | 0.0274 | 3.1460 | 0.8389 |
| G4 | 0.2 | N/A | N/A | 3.9470 | 0.6134 |
| G6 | 0.02 | 2302.206 | 361.058 | 0.3558 | 0.2080 |
| G6 | 0.1 | 3.0233 | 0.3855 | 2.7632 | 0.3646 |
| G6 | 0.15 | 0.2660 | 0.0190 | 4.5873 | 1.4219 |
| G6 | 0.2 | N/A | N/A | 15.3473 | 8.2928 |

For G4 dendrimer solutions with concentration $c > 0.1$, it is found that the $Q^*$ determined from the SANS model fitting is close to the $Q$ range relevant to the internal collective motion as shown in Figure 1. Therefore, in this case the value of $D_S^{Exp}(Q^*)$ cannot be entirely attributed to the hydrodynamic contribution. In addition it also contains the information from the internal collective motion. Nevertheless, the $D_S^{Exp}(Q)$ obtained from G4 dendrimer solutions are relatively featureless. In comparison to the characteristic variation of those of G6, their fluctuation is much less significant within the probed $Q$ range. Also the $D_S^{Exp}(Q)$ within the $Q$ range relevant to internal collective motion is seen to be significantly suppressed upon increasing $c$. Therefore although $\tau_{inter}$ and $\tau_{intra}$ for G4 dendrimer solution cannot be individually determined based on our phenomenological approach, it can be deduced that they are indeed within the same order of magnitude and their difference is steadily reduced upon increasing $c$ within the range of $0 < c < c^*$. This is consistent with the conclusion drew from the G6 dynamical data analysis.

## SIV. Small-Angle Neutron Scattering Experiment

Small-angle neutron scattering (SANS) measurements were respectively performed at the D22 SANS spectrometer at the ILL and at the EQ-SANS instrument at the Spallation Neutron Source (SNS) located at Oak Ridge National Laboratory (ORNL) using 60 Hz operation. For the SANS experiment at ILL, The wavelength of the incident neutron beam was chosen to be 6.0 Å, with a wavelength spread $\Delta\lambda/\lambda$ of 10%, to cover values of the scattering wave vector $Q$ ranging from 0.008 to 0.45 Å$^{-1}$. The measured intensity $I(Q)$ was corrected for detector background and sensitivity and for the scattering contribution from the empty cell and placed on an absolute scale using a direct beam measurement.

For the SANS experiment at SNS, Three configurations of sample-to-detector distance of 4 m, 2.5m, and 1.3m with two neutron wavelength bands were used to cover the $Q$ range of 0.008 Å$^{-1}$ $< Q < 0.6$ Å$^{-1}$ where $Q = (4\pi/\lambda) \sin(\theta/2)$ is the magnitude of the scattering vector, and $\theta$ is the scattering angle. The measured scattering intensity was corrected for detector sensitivity and the background from the empty cell and placed on an absolute scale using a calibrated standard.
All the SANS measurements at ILL and SNS were carried out using Hellma quartz cells of 1 mm path length at 20.0 °C ± 0.1 °C.

## SV. Small Angle Neutron Scattering Data Analysis

The zero-angle scattering of dendrimer solutions with number density $n$ at a certain D/H ratio of $\gamma$ can be expressed as

$$I_\gamma(0) = nP(0)S(0) \tag{S22}$$

where $n$ is the number density of dendrimer, $P(0)$ and $S(0)$ are the values of dendrimer form factor $P(Q)$ and inter-dendrimer structure factor $S(Q)$ at $Q = 0$. The value of $I_\gamma(0)$ can be obtained experimentally using the Guinier analysis.

The total scattering power of a single dendrimer is given by

$$P(0) = b_\gamma^2 = \left( \int_{v_{particle}} \Delta\rho(\vec{r}) d^3\vec{r} \right)^2 \tag{S23}$$

where $\Delta\rho(\vec{r})$ is the intra-dendrimer density profile along the radial direction. $b_\gamma$ is the summation of the bond scattering length of the whole dendrimer including the constituent polymer component and the density variation of the invasive water residing in the intra-dendrimer interior open space compared to the bulk solvent. Both of them contribute to the scattering power of the system. Explicitly, $b_\gamma$ can be expressed as

$$b_\gamma = b_{polymer} - \rho_\gamma v_{polymer} + \rho_\gamma (v_{water} \cdot h - 1) v_{cavity} \tag{S24}$$

where $b_{polymer}$ is the summation of the bond scattering length of the polymer components of dendrimer, $v_{polymer}$ is the summation of the volume of each constituent atom in the polymer component of dendrimer, $v_{water}$ is the molecular volume of water at its bulk state (30 Å$^3$), $h$ is the average number density of invasive water in the intra-dendrimer interior space, $v_{cavity}$ is the volume of the intra-dendrimer interior space where the density of the invasive water molecules is different from the bulk one, and

$$\rho_\gamma = \gamma \rho_{D_2O} + (1-\gamma) \rho_{H_2O} \tag{S25}$$

is the scattering length density (SLD) of the water at a certain D/H ratio $\gamma$. It is important to note that the first two terms present in the RHS of Eqn. (S24) give the coherent scattering contribution due to the SLD difference originating from the compositional difference between the polymer components of dendrimer and bulk water. The last term on the RHS side of Eqn. (S24) contributes to the coherent scattering via the density difference between the invasive water and its bulk state.

Assuming the isotope effect on inter-dendrimer spatial arrangement is negligible, for dendrimer solution with a certain concentration, the variation of scattering power for a single dendrimer due to the change of D/H ratio of the solvent from 100/0 to $\gamma$ can be expressed as

$$\frac{P_\gamma(0)}{P_{D/H=100/0}(0)} = \frac{I_\gamma(0)}{I_{D/H=100/0}(0)} \tag{S26}$$

Therefore, the variation of $b_\gamma$ can be expressed as

$$\frac{b_\gamma}{b_{D/H=100/0}} = \left(\frac{I_\gamma(0)}{I_{D/H=100/0}(0)}\right)^{1/2} = \frac{b_{polymer} - \rho_\gamma \left[v_{polymer} + (1-v_{water} \cdot h)v_{cavity}\right]}{b_{polymer} - \rho_{D/H=100/0} \left[v_{polymer} + (1-v_{water} \cdot h)v_{cavity}\right]} \tag{S27}$$

If we define

$$\theta \equiv \left(\frac{I_\gamma(0)}{I_{D/H=100/0}(0)}\right)^{1/2} \tag{S28}$$

One can obtain the average SLD of dendrimer

$$\langle\rho\rangle \equiv \frac{b_{polymer}}{v_{polymer} + (1-v_{water} \cdot h)v_{cavity}} \tag{S29}$$

It is found that $\theta$ and $\langle\rho\rangle$ can be respectively expressed as

$$\theta = \frac{\langle\rho\rangle - \rho_\gamma}{\langle\rho\rangle - \rho_{D_2O}} \tag{S30}$$

and

$$\langle\rho\rangle = \frac{\rho_\gamma - \rho_{D_2O}\theta}{1-\theta} \tag{S31}$$

Since the operation of division in Eqn. (S27) is carried out for dendrimer solutions with the same concentration but different contrasts, the effect that may potentially cause the shift of $I(0)$ due to the extrapolation, from the same low $Q$ point for different form factors at different concentrations, can therefore be fully eliminated. Because $\rho_\gamma$, $\rho_{D_2O}$ and $\theta$ are either known physical quantities or experimentally measurable, from the coherent scattering cross section $I(Q)$ obtained from the contrast variation SANS experiment, as given in Figures S5 and S6, one can evaluate the average SLD variation of dendrimer as a function of dendrimer weight fraction $c$. It is important to point out that, since $b_{polymer}$ is a constant, the magnitude and the evolution of $\langle\rho\rangle$ shows the variation of the volume of the total scatterers including both the polymeric part and the cavity region.

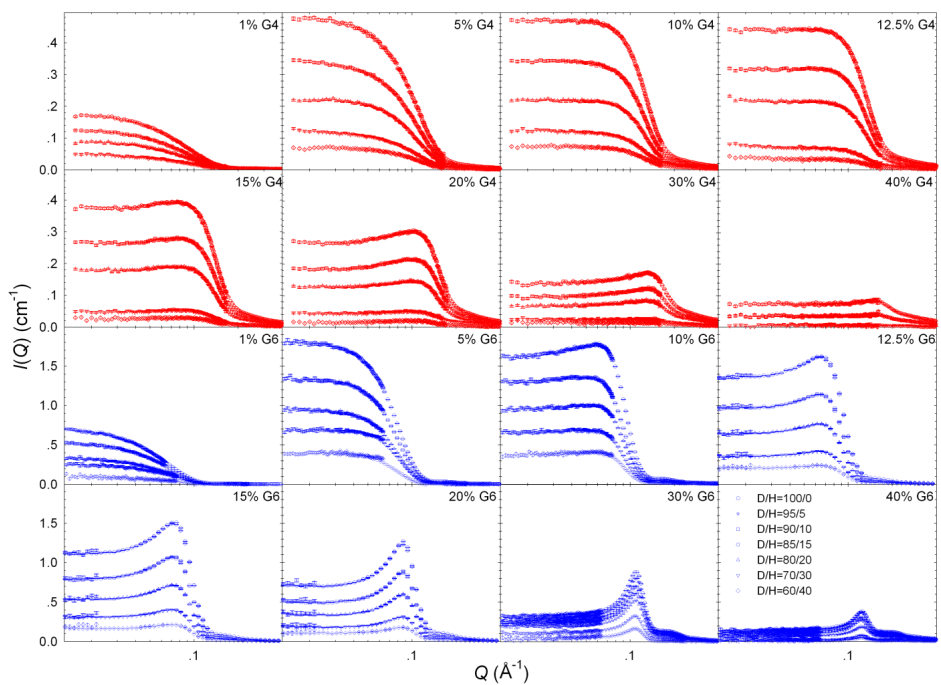

**FIG. S5.** (color online). The coherent scattering cross section $I(Q)$ obtained from G4 and G6 PAMAM dendrimer solutions with dendrimer weight fraction ranging from 0.01 to 0.4. The magnitude of the $I(Q)$ is seen to decrease continuously upon changing the D/H ratio from 100/0 to 60/40.

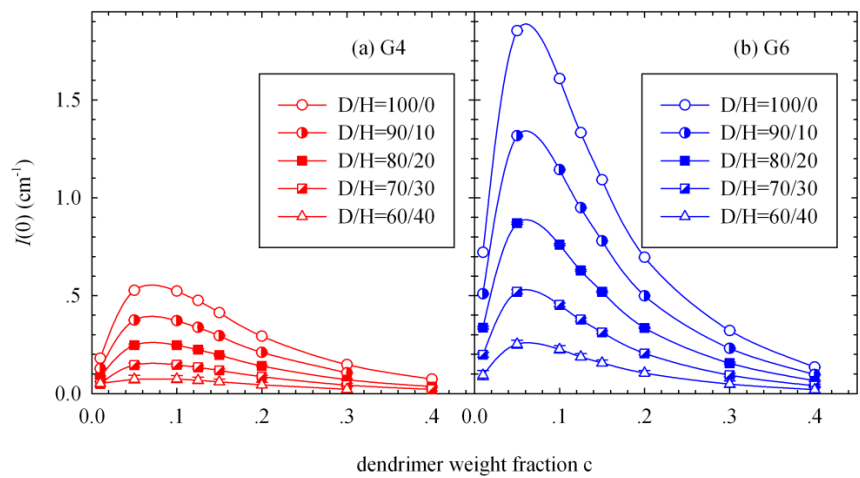

**FIG. S6.** (color online). The variation of $I(0)$ as a function of dendrimer weight fraction c obtained from the extrapolation of the $I(Q)$ presented in Figure S5.

### SVI. Behavior of $<\rho>$ when $c > c^*$

The evolution of $<\rho>$ beyond $c^*$ was also explored in this study. As shown in Figure S7, when $c > c^*$ the $<\rho>$ for G4 and G6 dendrimers dissolved in water in general exhibits no dependence on $c$ if the experimental errors are taken into consideration. Several factors could collectively contribute to this observation. First it is possible that the exceedingly significant steric crowding beyond $c^*$ could essentially contract most of the existing intra-dendrimer open space and therefore eliminate the generational dependence of $<\rho>$. Because of the decreasing local density fluctuation caused by the inter-particle congestion, [S14-S15] as shown in Figure S5, significant diminishing coherent scattering intensity is observed when $c > c^*$ even when the dendrimer is at full contrast against the deuterium oxide. With this constraint, it is difficult to explore any further detailed evolution of $<\rho>$ with improving statistics based on our current approach of contrast variation. Moreover, in our approach, the total scattering power of a single dendrimer is obtained by eliminating the inter-dendrimer interaction at different $c$. However, it has been indicated that [S2] the factorization approximation, [S16] which conveniently facilitates the determination of intra- and inter-particle correlation separately from the experimental coherent scattering cross section $I(Q)$, is no longer valid at higher concentration. When $c > c^*$, the mathematical expression of $<\rho>$ given by Eqn. (1) may no longer be valid since the $I(Q)$ collected from the highly concentrated dendrimer solutions cannot be expressed as a product of form factor $P(Q)$ and structure factor $S(Q)$. Therefore, while the dehydration of a single dendrimer can be clearly deduced from the variation of $<\rho>$ when $c < c^*$, beyond $c^*$ the evolution of $<\rho>$ cannot be considered as a reflection of the conformational evolution of a single dendrimer alone.

### SVII. Procedure of Atomistic Molecular Dynamics (MD) Simulations

Atomistic MD simulations were used to calculate the configuration of interfacial water presented in Figure 3. The details can be found in references [S17-S19].

The packing of these invasive water molecules can be visualized from the MD-calculated partial pair distribution functions $g_{oo}(r)$ of the constituent oxygen atoms. The correlation between the invasive water and its distance between the nearest polymer components of dendrimer is shown in Figure S8. A steady decrease in the height of the first peak of $g_{oo}(r)$, which contains the information of the local ordering, is observed once the distance between invasive water and the polymer is less than 5 Å. This suggests that the packing of invasive water is indeed looser than that of its bulk state. The coherent neutron scattering cross section is known to be the manifestation of the scattering length density (SLD) of the whole studied system. Therefore, from an experimental perspective, both polymer components of dendrimer and the invasive water indeed give rise to the coherent scattering contribution via both the compositional and density differences from the bulk water background respectively.

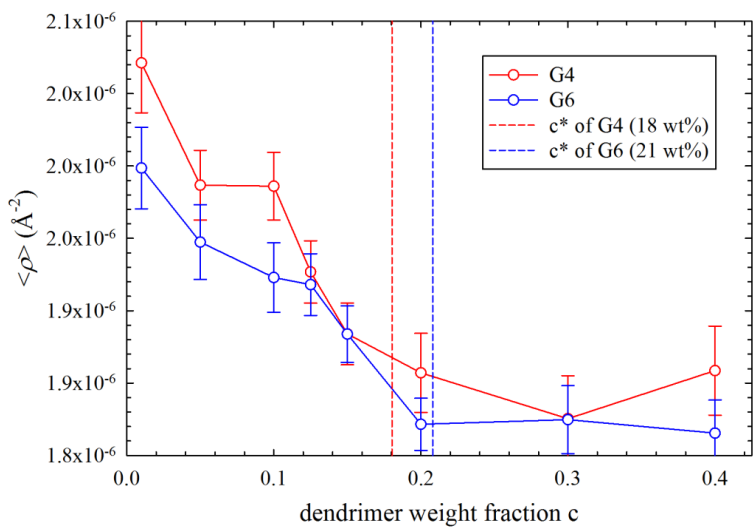

**FIG. S7.** (color online). The evolution of $<\rho>$ as a function of dendrimer weight fraction from $c = 0.02$ to $c = 0.4$.

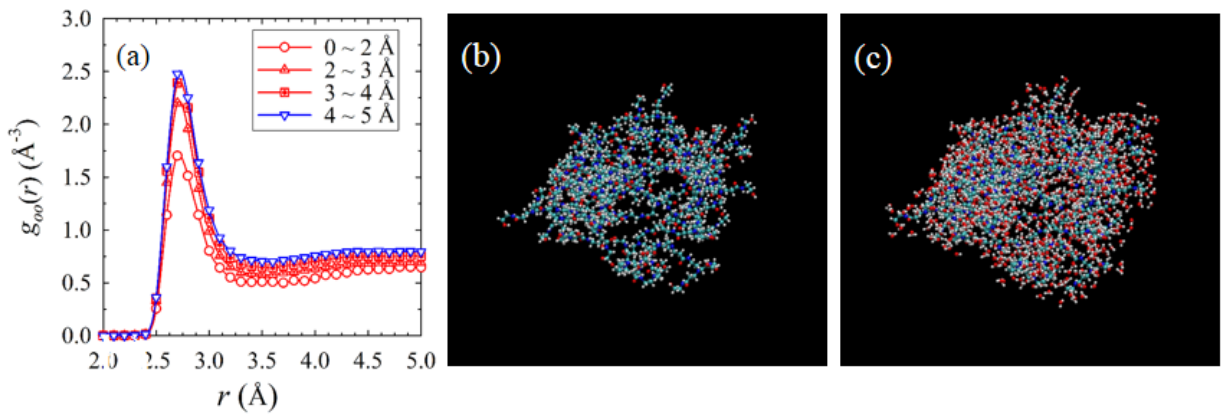

**FIG. S8** (a) The averaged oxygen-oxygen pair distribution function $g_{oo}(r)$ of TIP3P water around the vicinity of a PAMAM dendrimer. The snapshots of hydrocarbon components of a dendrimer (b) along with the surrounding invasive water (c) up to 5 Å.